\documentclass[12pt]{article}
\input{psfig.sty}
\input{epsf.sty}
\usepackage{graphicx}
\usepackage{amsmath,amssymb}
\usepackage{color}
\usepackage{epsfig}
\newcommand{\Define}{\stackrel{\triangle}{=}}
\newcommand{\inprob}{\stackrel{p}{\longrightarrow}}
\newtheorem{thm}{\bf Theorem}

\newtheorem{lem}{{Lemma}}

\begin{document}
\baselineskip 0.25in

\title{\LARGE Asymptotic Analysis of the Performance of LAS 
Algorithm for Large-MIMO Detection\thanks{This paper in part was presented 
in IEEE PIMRC'2008, Cannes, France, September 2008.}
\\
\vspace{-3mm}
}
\author{
Saif K. Mohammed, A. Chockalingam, and B. Sundar Rajan
\\
{\normalsize Department of ECE, Indian Institute of Science, 
Bangalore 560012, India \vspace{-2.00cm}}  \\
}
\date{}
\maketitle

\baselineskip 1.20pc

\baselineskip 2.00pc
\vspace{-8.5mm}
\begin{abstract}
\vspace{-4.5mm}
In our recent work, we reported an exhaustive study on the simulated bit 
error rate (BER) performance of a low-complexity likelihood ascent search 
(LAS) algorithm for detection in large multiple-input multiple-output (MIMO) 
systems with large number of antennas that achieve high spectral efficiencies. 
Though the algorithm was shown to achieve increasingly closer to near 
maximum-likelihood (ML) performance through simulations, no BER analysis 
was reported. Here, we extend our work on LAS and report an asymptotic BER 
analysis of the LAS algorithm in the large system limit, where 
$N_t,N_r \rightarrow \infty$ with $N_t=N_r$, where $N_t$ and $N_r$ are the
 number of transmit and receive antennas. We prove that the error 
performance of the LAS detector in V-BLAST with 4-QAM in i.i.d. Rayleigh 
fading converges to that of the ML detector as $N_t,N_r \rightarrow \infty$. 
\end{abstract}

\vspace{-8.0mm}
\baselineskip 1.725pc

\vspace{-7.5mm}
\section{Introduction}
\label{sec1}
\vspace{-8.5mm}
Multiple-input multiple-output (MIMO) systems that employ large number of 
transmit and receive antennas can offer very high spectral efficiencies
of the order of tens to hundreds of bps/Hz
\cite{tela99},\cite{paulraj}. Achieving near-optimal signal detection
at low complexities in such large-dimension systems has been a challenge. 
In our recent works, we have shown that certain algorithms from machine 
learning/artificial intelligence achieve near-optimal performance in 
large-MIMO systems that employ tens of transmit and receive antennas
using V-BLAST and non-orthogonal space-time block codes (STBC) \cite{bsr}
with tens to hundreds of dimensions in space and time, at low complexities.
Such algorithms include local neighborhood search based algorithms like 
a {\em likelihood ascent search} (LAS) algorithm \cite{jsac},\cite{stbc}
and a {\em reactive tabu search} (RTS) algorithm \cite{rts}, and algorithms
based on {\em probabilistic data association} (PDA) \cite{pda} and {\em
belief propagation} (BP) \cite{bp1},\cite{bp2}. Similar algorithms have 
been earlier reported in the context of multiuser detection 
\cite{las1}-\cite{bpmud2}. In \cite{jsac}-\cite{bp2}, through detailed 
simulations, we have shown that LAS and RTS algorithms achieve increasingly 
closer to maximum-likelihood (ML) performance and that PDA and BP algorithms 
achieve near maximum a posteriori probability (MAP) performance for increasing 
number of dimensions in large-MIMO systems. For e.g., in \cite{stbc}, the BER 
performance of the basic LAS algorithm (which uses a single symbol update 
based neighborhood definition) and its generalized version (which uses a 
multiple symbol update based neighborhood definition) has been exhaustively 
studied through simulations. However, BER performance analysis of the LAS 
algorithm for large-MIMO detection has not been reported. In this 
correspondence, we fill some of this gap by presenting an asymptotic 
BER analysis of the LAS algorithm in the large system limit, where 
$N_t,N_r \rightarrow \infty$ with $N_t=N_r$, where $N_t$ and $N_r$ denote 
the number of transmit and receive antennas, respectively. Asymptotic 
performance analysis of large systems in the context of multiuser 
detection and MIMO communication have been reported in the literature 
\cite{lma1}-\cite{lma6}, using random matrix theory (e.g., 
\cite{lma1},\cite{lma2}), replica method (e.g., 
\cite{lma3},\cite{lma4},\cite{lma5}), and free probability theory
(e.g., \cite{lma6}). We, in this correspondence, present an asymptotic BER 
analysis of the LAS algorithm in the large system limit. Specifically, 
we present an analytical proof that the error performance of the LAS 
detector for V-BLAST with 4-QAM in i.i.d. Rayleigh fading converges to 
that of the ML detector as $N_t,N_r \rightarrow \infty$ with $N_t=N_r$,
which is an analytical result that has not been reported so far.

The rest of this correspondence is organized as follows. The MIMO system 
model and LAS detection algorithm are summarized in Section \ref{sec2}. 
The asymptotic analysis of the LAS algorithm is presented in Section 
\ref{sec3}. Lengthy proofs of lemmas and theorems are moved to the 
appendices. Simulation results and discussions are presented in Section 
\ref{sec4}. Conclusions are given in Section \ref{sec5}.

\vspace{-5mm}
\section{System Model}
\label{sec2}
\vspace{-5mm}
Consider a V-BLAST system with $N_t$ transmit antennas and $N_r$ receive
antennas, $N_t\leq N_r$.  Let ${\bf x}_c \in {\mathbb C}^{N_t \times 1}$ 
denote\footnote{Vectors are denoted by boldface lowercase letters, and 
matrices are denoted 
by boldface uppercase letters. $[.]^T$ and $[.]^H$ 
denote transpose and conjugate transpose operations, respectively.
$||.||$ denotes Euclidean distance.
} 
the symbol vector transmitted, and 
${\bf H}_c \in {\mathbb C}^{N_r \times N_t}$ denote the channel matrix 
such that its $(i,j)$th entry $h_{i,j}$ is the complex channel gain from 
the $j$th transmit antenna to the $i$th receive antenna. Assuming rich 
scattering, we model the entries of ${\bf H}_c$ as i.i.d.
$\mathcal C \mathcal N(0,1)$.  Let ${\bf y}_c \in {\mathbb C}^{N_r \times 1}$ 
and ${\bf n}_c \in {\mathbb C}^{N_r \times 1}$ denote the received signal 
vector and the noise vector, respectively, at the receiver, where the entries 
of ${\bf n}_c$ are modeled as i.i.d $\mathcal C \mathcal N(0,\sigma^2)$. The
received signal vector can then be written as
\vspace{-3mm}
\begin{eqnarray}
\label{SystemModel}
{\bf y}_c & = & {\bf H}_c{\bf x}_c + {\bf n}_c.
\end{eqnarray}

\vspace{-7mm}
Let ${\bf y}_c$, ${\bf H}_c$, ${\bf x}_c$, and ${\bf n}_c$ be decomposed
into real and imaginary parts as follows:
\vspace{-2mm}
\begin{eqnarray}
\label{SystemModelDecompose}
{\bf y}_c \, = \, {\bf y}_I + j{\bf y}_Q, \,\,\,\,\,\,\,  {\bf x}_c \, = \, {\bf x}_I + j{\bf x}_Q,
\,\,\,\,\,\,\, 
{\bf n}_c \, = \, {\bf n}_I + j{\bf n}_Q, \,\,\,\,\,\,\, {\bf H}_c \, = \, {\bf H}_I + j{\bf H}_Q.
\end{eqnarray}

\vspace{-4mm}
Further, we define
${\bf H}_r \in {\mathbb R}^{2N_r \times 2N_t}$,
${\bf y}_r \in {\mathbb R}^{2N_r \times 1}$,
${\bf x}_r \in {\mathbb R}^{2N_t \times 1}$, and
${\bf n}_r \in {\mathbb R}^{2N_r \times 1}$ as

\vspace{-10mm}
\begin{eqnarray}
\label{SystemModelRealDef} 
{\bf H}_r \, = \, \left(\begin{array}{cc}{\bf H}_I \hspace{2mm} -{\bf H}_Q \\
{\bf H}_Q  \hspace{5mm} {\bf H}_I \end{array}\right), \,\,\,\,\, 
{\bf y}_r \, = \, [{\bf y}_I^T \hspace{2mm} {\bf y}_Q^T ]^T, \,\,\,\,\,
{\bf x}_r \, = \, [{\bf x}_I^T \hspace{2mm} {\bf x}_Q^T ]^T, \,\,\,\,\,
{\bf n}_r \, = \, [{\bf n}_I^T \hspace{2mm} {\bf n}_Q^T ]^T.
\end{eqnarray}

\vspace{-3mm}
Now, (\ref{SystemModel}) can be written as

\vspace{-15mm}
\begin{eqnarray}
\label{SystemModelReal}
{\bf y}_r & = & {\bf H}_r{\bf x}_r + {\bf n}_r.
\end{eqnarray}

\vspace{-5mm}
Henceforth, we shall work with the real-valued signal model of the
system in (\ref{SystemModelReal}). For notational simplicity,
we drop subscripts $r$ in (\ref{SystemModelReal}) and write

\vspace{-15mm}
\begin{eqnarray}
\label{SystemModelII}
{\bf y} & = & {\bf H} {\bf x} + {\bf n},
\end{eqnarray}
where {\small ${\bf H} = {\bf H}_r \in {\mathbb R}^{2N_r \times 2N_t}$},
{\small ${\bf y} = {\bf y}_r \in {\mathbb R}^{2N_r \times 1}$},
{\small ${\bf x} = {\bf x}_r \in {\mathbb R}^{2N_t \times 1}$},
{\small ${\bf n} = {\bf n}_r \in {\mathbb R}^{2N_r \times 1}$}.
In this real-valued system model, the real-part of the complex data
symbols will be mapped to {\small $[x_1,\cdots,x_{N_t}]$} and the
imaginary-part of these symbols will be mapped to
{\small $[x_{N_t+1},\cdots,x_{2N_t}]$}. For $M$-QAM,
{\small $[x_1,\cdots,x_{N_t}]$} can be viewed to be from an underlying
$M$-PAM signal set and so is {\small $[x_{N_t+1},\cdots,x_{2N_t}]$.}
Let $\mathbb A_i$ denote the $M$-PAM signal set from which $x_i$ takes
values, $i=1,2,\cdots,2N_t$; e.g., for 4-QAM,
${\mathbb A}_i=\{1,-1\}$ for $i=1,2,\cdots,2N_t$. Now, define a
$2N_t$-dimensional signal space $\mathbb S$ to be the Cartesian product
of $\mathbb A_1$ to $\mathbb A_{2N_t}$. The ML solution vector,
${\bf d}_{ML}$, is given by

\vspace{-10mm}
\begin{eqnarray}
\label{MLdetection}
{\bf d}_{ML} & = & {\mbox{arg min}\atop{{\bf d} \in {\mathbb S}}}
\hspace{1mm}  \Vert {\bf y} - {\bf H}{\bf d} \Vert ^2  
\,\,\,\, = \,\,\,\, {\mbox{arg min}\atop{{\bf d} \in {\mathbb S}}}
\hspace{1mm}\left({\bf d}^T {\bf H}^T{\bf H}{\bf d} - 2{\bf y}^T{\bf H}{\bf d}\right).
\end{eqnarray}

\vspace{-4mm}
In the following subsection, we summarize the low-complexity LAS 
algorithm, using a neighborhood definition based on 1-symbol 
updates, presented in \cite{stbc} for large-MIMO detection for $M$-QAM. 
The channel matrix {\bf H} is assumed to be known perfectly at the receiver.

\vspace{-3mm}
\subsection{LAS Algorithm for Large-MIMO Detection} 
\label{sec2a}
\vspace{-4mm}
The LAS algorithm starts with an initial vector ${\bf d}^{(0)}$,
given by ${\bf d}^{(0)} =  {\bf B}{\bf y}$, where ${\bf B}$ is
the initial solution filter, which can be a matched filter (MF) or
zero-forcing (ZF) filter or MMSE filter. The index $m$ in ${\bf d}^{(m)}$
denotes the iteration number in a given search stage. The ML cost function
after the $k$th iteration in a given search stage is given by
\vspace{-1mm}
\begin{eqnarray}
\label{Ck}
C^{(k)} & = & {\bf d}^{(k)^T} {\bf H}^T{\bf H} {\bf d}^{(k)}-2{\bf y}^T{\bf H}{\bf d}^{(k)}.
\end{eqnarray}

\vspace{-6mm}
The ${\bf d}$ vector is updated from $k$th to $(k+1)$th iteration
by updating one symbol, say, the $p$th symbol, as

\vspace{-17.0mm}
\begin{eqnarray}
\label{UpdateK}
{\bf d}^{(k+1)}  & = & {\bf d}^{(k)} + \lambda_p^{(k)} {\bf e}_p,
\vspace{-1mm}
\end{eqnarray}
where ${\bf e}_p$ denotes the unit vector with its $p$th entry only as one,
and all other entries as zero. Since ${\bf d}^{(k)}$ and ${\bf d}^{(k+1)}$
should belong to $\mathbb S$, $\lambda_p^{(k)}$ can take only certain integer 
values. For example, for 16-QAM, ${\mathbb A}_p = \{ -3, -1, 1, 3\}\big)$, 
and $\lambda_p^{(k)}$ can take values only from $\{-6,-4,-2,0,2,4,6\}$. 
Using (\ref{Ck}) and (\ref{UpdateK}), and defining a matrix ${\bf G}$ as
\vspace{-2mm}
\begin{eqnarray}
{\bf G} & \Define & {\bf H}^{T}{\bf H},
\end{eqnarray}

\vspace{-4mm}
we can write the cost difference
{\small $C^{(k+1)} - C^{(k)}$} as
\vspace{-1mm}
\begin{eqnarray}
\label{Ck+1MinusCk}
\hspace{-0mm} \nonumber
\mathcal F(l_p^{(k)}) & \Define & C^{(k+1)} - C^{(k)} \,\,\, = \,\,\, l_p^{(k)^2}a_p - 2l_p^{(k)} \vert z_p^{(k)} \vert,
\end{eqnarray}

\vspace{-4mm}
where $z^{(k)}_p$ is the $p$th entry of the ${\bf z}^{(k)}$ vector
given by {\small ${\bf z}^{(k)}={\bf H}^T({\bf y}-{\bf H}{\bf d}^{(k)})$},
$a_p \Define \left({\bf G}\right)_{p,p}$ is the $(p,p)$th entry of the
${\bf G}$ matrix, and $l_p^{(k)} = \vert \lambda_p^{(k)} \vert$. The value
of $l_p^{(k)}$ which gives the largest descent in the cost function from
the $k$th to the $(k+1)$th iteration (when symbol $p$ is updated) is
obtained as

\vspace{-8mm}
{\small
\begin{eqnarray}
\label{Optdp}
l_{p,opt}^{(k)} & = & 2 \left\lfloor \frac { \vert z_p^{(k)} \vert } { 2 a_p} \right\rceil,
\end{eqnarray}
}

\vspace{-7mm}
where $\lfloor . \rceil$ denotes the rounding operation.
If $d_p^{(k)}$ were updated using $l_{p,opt}^{(k)}$, it is
possible that the updated value does not belong to ${\mathbb A}_p$.
To avoid this, we adjust $l_{p,opt}^{(k)}$ so that the updated value
of $d_p^{(k)}$ belongs to ${\mathbb A}_p$.  Let

\vspace{-14mm}
\begin{eqnarray}
s & = & {\mbox{arg min}\atop p} \,\,\,\mathcal F(l_{p,opt}^{(k)}).
\end{eqnarray}

\vspace{-3mm}
If {\small $\mathcal F(l_{s,opt}^{(k)}) < 0$}, the update for the
$(k+1)$th iteration is
\begin{eqnarray}
\label{AlgoUpdateK}
{\bf d}^{(k+1)} & = & {\bf d}^{(k)} + l_{s,opt}^{(k)}\,\mbox{sgn}(z_s^{(k)})\,{\bf e}_s \\
{\bf z}^{(k+1)} & = & {\bf z}^{(k)} - l_{s,opt}^{(k)}\,\mbox{sgn}(z_s^{(k)})\,{\bf g}_s,
\label{eqx}
\end{eqnarray}

\vspace{-6mm}
where ${\bf g}_s$ is the $s$th column of ${\bf G}$. If
$\mathcal F(l_{s,opt}^{(k)}) \geq 0$, then the search terminates, and
${\bf d}^{(k)}$ is declared as the detected data vector.

\vspace{-4mm}
\section{Asymptotic Analysis of LAS Algorithm}
\label{sec3}
\vspace{-5mm}
In this section, we prove the asymptotic convergence of the error
probability of the LAS detector to that of the ML detector
for $N_t,N_r \rightarrow \infty$ with $N_t=N_r$ in V-BLAST. Consider 4-QAM, 
i.e., ${\mathbb S} \in \{+1,-1\}^{2N_t}$, and let $N_t=N_r$.
An $n$-symbol update on a data vector ${\bf d} \in {\mathbb S}$ transforms
${\bf d}$ to $({\bf d}-{\Delta {\bf d}_n})$ such that
$({\bf d}-{\Delta {\bf d}_n}) \in {\mathbb S}$. Further,
$({\bf d}-{\Delta {\bf d}_n})$ is obtained by changing $n$ symbols
in ${\bf d}$ at distinct indices given by the $n$-tuple ${\bf u}_n$
{\small $\Define$ $(i_1,i_2,\cdots,i_n)$, $1 \leq i_j \leq 2N_t,
\forall j=1,\cdots,n$} and
$i_j\neq i_k \, \mbox{for} \, j\neq k$. 
Therefore, we can write $\Delta {\bf d}_n$ as
\vspace{-0mm}
\begin{equation}
\label{l1p1}
{\Delta {\bf d}_n} = \sum_{k=1}^{n}  2d_{i_{k}} {\bf e}_{i_{k}},
\end{equation}

\vspace{-2mm}
where $d_{i_{k}}$ is the $i_k$th element of ${\bf d}$.
Let ${\mathbb L}_n \subseteq {\mathbb S}$ denote the set of data vectors
such that for any ${\bf d} \in {\mathbb L}_n$, if a $n$-symbol update
is performed on ${\bf d}$ resulting in a vector $({\bf d}-\Delta {\bf d}_n)$,
then
$||{\bf y}-{\bf H}({\bf d}-\Delta{\bf d}_n)||\geq||{\bf y}-{\bf H}{\bf d}||$.
Our main result in this section is Theorem \ref{thm2}. To prove 
Theorem \ref{thm2}, we need the following Lemmas \ref{lem1} to \ref{lem5}, 
Slutsky's theorem \cite{basu}, and Theorem \ref{thm1}. 

\vspace{-6mm}
\begin{lem}
\label{lem1}
Let ${\bf d} \in {\mathbb S}$. Then, ${\bf d} \in {\mathbb L}_n$
if and only if, for any $n$-update on ${\bf d}$, $n \in [1,2\cdots,2N_t]$,

\vspace{-13mm}
\begin{eqnarray}
\label{l1p2}
\big({\bf y} - {\bf H}{\bf d} + \frac {1}{2} {\bf H}{\Delta {\bf d}_n}\big)^T\big({\bf H}{\Delta {\bf d}_n}\big) & \geq & 0.
\end{eqnarray}
\end{lem}

\vspace{-4mm}
{\em Proof:}
By definition, if 
${\bf d} \in {\mathbb L}_n$, then no $n$-symbol update can result in a 
reduction in the ML cost function. Using this, we can write 
\begin{eqnarray}
\label{pf1p1}
\Vert {\bf y} -  {\bf H}({\bf d} - {\Delta {\bf d}_n}) \Vert^2 & \geq & \Vert {\bf y} -  {\bf H}{\bf d} \Vert^2.
\end{eqnarray}
Simplifying (\ref {pf1p1}), we get (\ref {l1p2}). Since the choice 
of the indices in ${\bf u}_n$ is arbitrary, the lemma holds true for all 
possible $n$-tuples of distinct indices. For the converse, if ${\bf d}$
satisfies (\ref{l1p2}) for all possible ${\bf u}_n$ for a given $n$,
then, since (\ref{l1p2}) and (\ref{pf1p1}) are equivalent, ${\bf d}$
also satisfies (\ref{pf1p1}) for all possible ${\bf u}_n$. This implies 
that ${\bf d} \in {\mathbb L}_n$. 
$\square$

\vspace{-3mm}
If ${\bf d} \in {\mathbb L}_1$, then using Lemma \ref{lem1} and 
(\ref{SystemModelII}),
we can write
\vspace{-1mm}
\begin{equation}
\label{l1p5}
\hspace{-1mm}
\big({\bf n} + {\bf H}({\bf x} - {\bf d}) + {\bf h}_{p}{d}_{p} \big)^T \big({\bf h}_{p}{d}_{p}) \geq 0, \,\, \forall p = 1,\cdots,2N_t, \hspace{-2mm}
\end{equation}

\vspace{-5mm}
where ${\bf h}_p$ is the $p$th column of {\bf H}.

\vspace{-3mm}
\begin{lem}
\label{lem2}
Assuming uniqueness of the ML vector ${\bf d}_{ML}$ in (\ref{MLdetection}), 
a symbol vector ${\bf d} \in {\mathbb S}$ is the ML vector if and only if 
the noise vector ${\bf n}$ satisfies the following set of equations
\begin{eqnarray}
\label{MLCond}
\Big({\bf n} + {\bf H}({\bf x} - {\bf d}) + \Big(\sum_{j = 1}^{n} {\bf h}_{i_{j}}{d}_{i_{j}}\Big)\Big)^T \Big(\sum_{j = 1}^{n} {\bf h}_{i_{j}}{d}_{i_{j}}\Big) \geq 0,
\end{eqnarray}

\vspace{-4mm}
$\forall \, n=1,\cdots,2N_t$, and for all possible $n$-tuples
$(i_1,\cdots,i_n)$ for each $n$.
\end{lem}
\vspace{-4mm}
{\em Proof:}
If ${\bf d}$ is the unique ML vector, then from the definition of the ML 
criterion in (\ref{MLdetection}), it must be true that any $n$-update on 
${\bf d}$ will not 
result in any decrease in the ML cost function. Therefore, 
${\bf d} \in {\mathbb L}_n$, $\forall$ $n = 1,2 \cdots,2N_t$. Hence, by 
Lemma \ref{lem1}, it must be true that ${\bf d}$ satisfies (\ref{l1p2}) 
for all $n = 1,2,\cdots,2N_t$ and for all possible ${\bf u}_n$ for each $n$. 
Substituting ${\bf y} = {\bf H}{\bf x} + {\bf n}$ in (\ref{l1p2}), we 
get (\ref{MLCond}). This proves the direct result. To prove the converse,
let the noise vector ${\bf n}$ satisfy (\ref{MLCond}) for some vector 
${\bf d}$. Since ${\bf y} = {\bf H}{\bf x} + {\bf n}$, the conditions in 
(\ref{MLCond}) imply the conditions in (\ref{l1p2}) for all 
$n=1,2,\cdots,2N_t$ and for all possible ${\bf u}_n$ for each $n$. 
Therefore, by Lemma \ref{lem1}, ${\bf d} \in {\mathbb L}_n$ for all 
$n=1,2,\cdots,2N_t$, which then implies that ${\bf d}$ indeed is the 
ML vector.
$\square$

\vspace{-1mm}
{\em Definition:} For each ${\bf d} \in {\mathbb S}$ and for each
integer $m$, $1 \leq m \leq 2N_t$, we associate the set of vectors 
${\cal R}_{{\bf d}^m} = \Big\{{\bf v} \,|\, {\bf v} \in {\mathbb R}^{2N_t} 
\, \mbox{and} \,
\big({\bf v} + {\bf H}({\bf x} - {\bf d}) + \big(\sum_{j = 1}^{n} {\bf h}_{i_{j}}{d}_{i_{j}}\big)\big)^T \big(\sum_{j = 1}^{n} {\bf h}_{i_{j}}{d}_{i_{j}}\big) \geq 0, \, \forall \, n=1,\cdots,m,
\mbox{and for all possible $n$-tuples} \, (i_1,\cdots,i_n) \, \mbox{for 
each} \, n \Big\}$, and 
define ${\cal R}_{\bf d} \Define {\cal R}_{{\bf d}^{2N_t}}$.

\vspace{-4mm}
\begin{lem}
\label{lem3}
If the noise vector ${\bf n} \in {\cal R}_{\bf d}$, then ${\bf d}$ is 
the ML vector.
Let ${\bf d}_i, {\bf d}_j \in {\mathbb S}$ and ${\bf d}_i \neq {\bf d}_j$.
Then ${\cal R}_{{\bf d}_i}$ and ${\cal R}_{{\bf d}_j}$ are disjoint.
\end{lem}

\vspace{-5mm}
{\em Proof:}
From Lemma \ref{lem2} and the definition of ${\cal R}_{\bf d}$, it is 
clear that ${\bf d}$ is the ML vector if and only if 
${\bf n} \in {\cal R}_{\bf d}$. 
The disjointness of ${\cal R}_{{\bf d}_i}$ and 
${\cal R}_{{\bf d}_j}$, $i\neq j$, can be shown by contradiction. If 
${\cal R}_{{\bf d}_i}$ and ${\cal R}_{{\bf d}_j}$ are not disjoint, 
then there exists some vector ${\bf v}$ belonging to both  
${\cal R}_{{\bf d}_i}$ and ${\cal R}_{{\bf d}_j}$. If ${\bf v}$ were 
to be the noise vector ${\bf n}$, then, ${\bf v}$ would satisfy the set 
of equations in (\ref{MLCond}) for both ${\bf d}={\bf d}_i$ and 
${\bf d}= {\bf d}_j$, since 
${\bf v}$ belongs to both ${\cal R}_{{\bf d}_i}$ and ${\cal R}_{{\bf d}_j}$,
This, by Lemma \ref{lem2}, implies that
both ${\bf d}_i$ and ${\bf d}_j$ are ML vectors, which is a 
contradiction because of the uniqueness of the ML vector.
$\square$

\vspace{-5mm}
\begin{lem}
\label{lem4}
Let ${\bf h} \in {\mathbb R}^{2N_t}$ be a random vector with i.i.d 
entries distributed as ${\cal N}(0,0.5)$. Let $\{{\bf h}_i\},i=1,2,\cdots,m$ 
be a set of vectors, with each ${\bf h}_i \in {\mathbb R}^{2N_t}$ and having 
i.i.d entries distributed as ${\cal N}(0,0.5)$, 
${\mathbb E}[{\bf h}_i{\bf h}_j^T]=0$ for $i\neq j$, and 
${\mathbb E}[{\bf h}{\bf h}_j^T]=0$ for $j=1,\cdots,m$.  Then 
\begin{eqnarray}
\label{l34St}
\lim_{_{N_t \rightarrow \infty}} \frac {\sum_{k=1}^{m} {\bf h}^T{\bf h}_k} {m N_t} & = & 0.
\end{eqnarray}
\end{lem}

\vspace{-5mm}
{\em Proof:}
Let ${\tilde {\bf h}} \Define \frac{1}{\sqrt{m}}\sum_{k=1}^{m} {\bf h}_k$.
Then, ${\tilde {\bf h}} \sim {\cal N}({\bf 0},\frac{\bf I}{2})$.
Therefore, we have
\vspace{-1mm}
\begin{eqnarray}
\label{l34St1}
\lim_{_{N_t \rightarrow \infty}} \frac {\sum_{k=1}^{m} {\bf h}^T{\bf h}_k} {m N_t}  & = & \lim_{_{N_t \rightarrow \infty}} \frac {{\bf h}^T{\tilde {\bf h}}} { \sqrt{m} N_t}.
\end{eqnarray}
We can write
\begin{eqnarray}
\label{l34St2}
\lim_{_{N_t \rightarrow \infty}} \frac {{\bf h}^T{\tilde {\bf h}}}{N_t} & = & 
\lim_{_{N_t \rightarrow \infty}} 
\frac {\sum_{k=1}^{2N_t} h_{k} {\tilde h}_{k}}{N_t},
\end{eqnarray}
where $h_{k}$ and ${\tilde h}_{k}$ are the $k$th elements of
${\bf h}$ and ${\tilde {\bf h}}$, respectively. The r.v's 
$h_k{\tilde h}_k, k=1,\cdots,2N_t$ are i.i.d with mean zero. 
From the strong law of large numbers \cite{basu}, it follows that
$\lim_{_{N_t \rightarrow \infty}} \sum_{k=1}^{2N_t} \frac{h_{k} {\tilde h}_{k}}{2N_t}=0$.
Using this in (\ref{l34St1}) completes the proof.
$\square$

\vspace{-2mm}
Before we present the next lemma, we present the Slutsky's theorem on 
convergence of random variables, which is used to prove Lemma \ref{lem5} 
and Theorem \ref{thm1}.

\vspace{-2mm}
{\em Slutsky's Theorem \cite{basu}:}
Let $\{ {\bf X}_m \}$ and $\{ {\bf Y}_m \}$ be sequences of random variables.
If $\{ {\bf X}_m \}$ converges in distribution to a random variable
${\bf X}$, and $\{{\bf Y}_m\}$ converges in probability to a constant
$c$, then it is true that
$i)$ $\{{\bf X}_m+{\bf Y}_m\}$ converges in distribution to ${\bf X}+c$, 
$ii)$ $\{{\bf X}_m {\bf Y}_m\}$ converges in distribution to $c{\bf X}$, 
and $iii)$ $\left\{\frac{{\bf X}_m}{{\bf Y}_m}\right\}$ converges in 
distribution to $\frac{{\bf X}}{c}$.

\vspace{-4mm}
\begin{lem}
\label{lem5}
For a given ${\bf u}_n$ and a given ${\bf d} \in {\mathbb S}$, 
define a r.v $z_{{\bf u}_n,{\bf d}}$ as
\vspace{-1mm}
\begin{eqnarray}
\label{l4St}
z_{{\bf u}_n,{\bf d}} & \Define & \frac {\sum_{k=1}^{n} \sum_{j=k+1}^{n} {\bf h}_{i_j}^T{\bf h}_{i_k} d_{i_j}d_{i_k}} { \sum_{j=1}^{n} \Vert {\bf h}_{i_j}\Vert^2},
\end{eqnarray}
where $i_j \in {\bf u}_n$, $j = 1,\cdots,n$. 
For any ${\bf u}_n$ and any ${\bf d} \in {\mathbb S}$, $z_{{\bf u}_n,{\bf d}}$ 
converges to zero in probability as $N_t\rightarrow \infty$, i.e., 
$z_{{\bf u}_n,{\bf d}} \inprob 0$
as $N_t\rightarrow \infty$, $\forall \, n=2,3,\cdots,2N_t$.
\end{lem}

\vspace{-4mm}
{\em Proof:} 
Proof of this Lemma is given in Appendix A.
$\square$

In Fig. \ref{fig0}, we plot the simulated pdf of $z_{{\bf u}_n,{\bf d}}$ for
$n=2N_t$ for different values of $N_t=N_r$ for a certain ${\bf u}_n$ and 
${\bf d}$ (the pdf was observed to be same for different ${\bf u}_n$ and
${\bf d}$). We observe that with
increasing $N_t=N_r$, the pdf of $z_{{\bf u}_n,{\bf d}}$ tends towards 
the Dirac 
delta function at zero. This implies that $z_{{\bf u}_n,{\bf d}}$ tends to
zero in distribution, and hence in probability, for large $N_t=N_r$, which 
is formally proved in Lemma \ref{lem5}.

\vspace{-4mm}
\begin{thm}
\label{thm1}
Let ${\bf d} \in {\mathbb S}$ and ${\bf n} \in {\cal R}_{{\bf d}^{1}}$.
Then ${\bf n} \in {\cal R}_{\bf d}$ in probability as
$N_t \rightarrow \infty$, i.e., for any $\delta$, $0 \leq \delta \leq 1$,
there exists an integer $N(\delta)$ such that for $N_t > N(\delta)$,
$p({\bf n} \in {\cal R}_{\bf d}) > 1-\delta$.
\end{thm}

\vspace{-4mm}
{\em Proof:} 
Proof of this theorem is given in Appendix B.
$\square$

\vspace{-4mm}
\begin{thm} 
\label{thm2} 
The data vector/bit error probability of the LAS detector converges
to that of the ML detector as $N_t,N_r \rightarrow \infty$ with
$N_t=N_r$.
\end{thm}

\vspace{-4mm}
{\em Proof:}
Let ${\bf d}_{LAS}$ be the final output symbol vector of the 
LAS algorithm given ${\bf x}$, ${\bf H}$ and ${\bf n}$. The algorithm
terminates if and only if no 1-update results in any further decrease of
the cost function. This implies that for the given ${\bf x}$, ${\bf H}$
and ${\bf n}$, ${\bf d}_{LAS} \in {\mathbb L}_1$, and therefore it must
be true that ${\bf n}$ satisfies (\ref{l1p5}) with ${\bf d}$ replaced by
${\bf d}_{LAS}$. These set of equations are the same which define
the region ${\cal R}_{{\bf d}^1}$. Therefore, replacing ${\bf d}$ by
${\bf d}_{LAS}$, we can equivalently claim that
${\bf n} \in {\cal R}_{{\bf d}_{LAS}^1}$. Using Theorem \ref{thm1}, we 
can further claim that asymptotically as $N_t \rightarrow \infty$,
${\bf n} \in {\cal R}_{{\bf d}_{LAS}}$ in probability. 
From Lemma \ref{lem3},
we know that if ${\bf n} \in {\cal R}_{d_{LAS}}$, then ${{\bf d}_{LAS}}$
is indeed the ML vector for the given ${\bf x}$, ${\bf H}$ and ${\bf n}$.
Therefore, we can state that asymptotically as $N_t \rightarrow \infty$,
${{\bf d}_{LAS}}$ is indeed the ML vector in probability.
That is, for any $\delta$, $0 \leq \delta \leq 1$, there exists an
integer $N(\delta)$ such that for $N_t \geq N(\delta)$
\begin{eqnarray}
\label{dlas_mlp}
P({{\bf d}_{LAS}} \,\, \mbox{is the ML vector}) & > & (1 - \delta).
\end{eqnarray}
Therefore, we can write that for $N_t \geq N(\delta)$

\vspace{-14mm}
\begin{eqnarray}
\label{dlas_mlp2}
\hspace{-2mm}
\nonumber
P_{LAS}(error) &\hspace{-2mm} = & \hspace{-2mm} P({\bf d}_{LAS} \neq {\bf x}) 
\,\,\, = \,\,\, P({\bf d}_{LAS} \neq {\bf x} \, | \, {\bf d}_{LAS} = \mbox{ML vector}) P({\bf d}_{LAS}  = \mbox{ML vector}) \nonumber \\
& & \hspace{20mm} + \, P({\bf d}_{LAS} \hspace{-0.25mm} \neq \hspace{-0.25mm} {\bf x} \, | \, {\bf d}_{LAS} \hspace{-0.0mm} \neq \mbox{ML vector}) P({\bf d}_{LAS} \neq \mbox{ML vector}).
\end{eqnarray}
From (\ref{dlas_mlp}), we have
$P({\bf d}_{LAS} \, \neq \, \mbox{ML vector}) \leq \delta$. Also,
$P({\bf d}_{LAS} \neq {\bf x} \, | \, {\bf d}_{LAS} = \mbox{ML vector})$
is the probability of error for the ML detector, which we denote by
$P_{ML}(error)$. Using these, we can bound the probability of
error for the LAS detector as
\vspace{-1mm}
\begin{eqnarray}
\hspace{-1mm}
\label{dlas_mlp3}
P_{LAS}(error) \,\, \leq \,\, P_{ML}(error) 
+ \, \delta \, P({\bf d}_{LAS} \neq {\bf x} \, | \, {\bf d}_{LAS} \neq \mbox{ML vector}) 
\,\, \leq \,\, {\bf P}_{ML}(error) + \delta.
\end{eqnarray}
Since $\delta$ can be arbitrarily small, we can conclude from
(\ref{dlas_mlp3}) that indeed as $N_t \rightarrow \infty$, the symbol
vector error probability of the LAS detector converges to that of
the ML detector. This proof can be adapted to show that apart
from the symbol vector error probability, the bit error probability of
the LAS detector also converges to that of the ML detector. The proof
for the bit error probability convergence is along the same lines as
(\ref{dlas_mlp2}) and (\ref{dlas_mlp3}), except that instead of defining
the error event as ${\bf d}_{LAS} \neq {\bf x}$, we define error events
for each bit. For example, for the $p$th bit, the error event is defined
as ${d_p}_{LAS} \neq {x_p}$. $\square$

\vspace{-4mm}
\section{Simulation Results and Discussions} 
\label{sec4}
\vspace{-5mm}
In Fig. \ref{fig1} we show the simulated BER performance of the LAS 
detector for V-BLAST with 4-QAM and MMSE initial vector for increasing 
$N_t=N_r$. Since an analytical expression for ML performance in the large 
MIMO system limit is not available and simulating the ML performance
for large dimensions involves prohibitively high complexity, we plot 
the SISO AWGN performance as a lower bound for comparison. It can be 
seen that for 
increasing $N_t=N_r$, the BER 
performance of the LAS detector approaches the SISO AWGN performance at 
high SNRs. Figure \ref{fig2} shows the average SNR required to achieve a 
BER of $10^{-3}$ for increasing $N_t=N_r$ and 4-QAM. It can be seen that, 
for large $N_t=N_r$, the required SNR gets increasingly closer to that 
required in SISO AWGN for increasing $N_t=N_r$. A similar behavior can
be observed in Fig. \ref{fig3} for 16-QAM as well. 
In Figs. \ref{fig2} and \ref{fig3}, we also see that there is an initial 
degradation in performance for increasing number of antennas (for $N_t<10$). 
This shows that the LAS detector is suboptimal for small systems with small 
number of antennas\footnote{We do not have a theoretical explanation for
this small system behavior of the LAS detector, whereas we are able to
prove its asymptotic large system behavior.}, and becomes optimal in the 
large system limit (as proved 
in the previous section). LAS detector achieves close to large system limit 
performance in systems with large number of dimensions (e.g., hundreds of 
dimensions in Figs. \ref{fig2} and \ref{fig3}). Such large number of 
dimensions need not be realized in spatial dimension alone, as in V-BLAST. 
As shown in \cite{stbc}, exploiting time dimension in addition to 
space dimension, large non-orthogonal STBC MIMO systems can render large 
dimensions with less number of transmit antennas that can be implemented in 
practice. A $16\times 16$ non-orthogonal STBC from cyclic division algebra
\cite{bsr} with complex data symbols has 512 real dimensions; with 64-QAM 
and rate-3/4 turbo code, this STBC achieves a spectral efficiency of 
72 bps/Hz. In \cite{stbc}, LAS algorithm has been shown to achieve 
near-capacity performance in $16\times 16$ STBC MIMO systems even in the
presence of spatial correlation and with estimated channel matrix. 
Further, considering that NTT DoCoMo has demonstrated a $12\times 12$ 
V-BLAST  MIMO system operating at 5 Gbps at a spectral efficiency of 
50 bps/Hz at 10 Km/hr mobile speeds \cite{docomo}, the availability of 
low-complexity large-MIMO detection algorithms like the LAS algorithm
analyzed in this correspondence can motivate the adoption of $16\times 16$ 
and $24\times 24$ MIMO systems operating at spectral efficiencies in excess 
of 50 bps/Hz in emerging wireless standards like IEEE 802.11 VHT 
and IEEE 802.16/LTE-A. 
 
\vspace{-4mm}
\section{Conclusions}
\label{sec5}
\vspace{-5mm}
We conclude with the following two remarks: $i)$ The derivation of analytical 
BER expressions for the ML performance in the large MIMO system limit for 
different signal sets is an open problem. Since large MIMO systems can be 
viable in practice due to the availability of low-complexity detectors like 
the LAS detector, analytical BER expressions for the ML performance in the 
large MIMO system limit would be quite useful as a benchmark for comparing 
the performance of practical detectors in large-MIMO systems. The statistical 
mechanics approach employed in \cite{lma3} for large CDMA system BER analysis 
can be investigated for such an analysis.  $ii)$ While we are able to prove 
the asymptotic convergence of LAS performance to ML performance for 4-QAM 
here, our simulation results for higher order QAM (e.g., 16-QAM; see Fig.  
\ref{fig3}) show similar behavioral trend like that for 4-QAM. Consequently, 
we conjecture that such a convergence holds for general $M$-QAM and an 
analytical proof to show this can be attempted as an extension to this work.

\vspace{-4mm}
\section*{Appendix A: Proof of Lemma \ref{lem5}}
\vspace{-5mm}
We present the proof of Lemma \ref{lem5} in this appendix.
The proof is by mathematical induction on $n$.
{\em Base Case:} For $n=2$, we have to show that
\begin{eqnarray}
\label{l4St1}
d_p \, d_q \, \frac { {\bf h}_p^T{\bf h}_q } { \Vert {\bf h}_p \Vert^2 + \Vert {\bf h}_q \Vert^2 } \inprob 0 \,\,\, \mbox{as} \,\, {N_t \rightarrow \infty}, \,\forall\, p,q = 1,2,\cdots,2N_t, \,\, p \ne q.
\end{eqnarray}
We can write the random variable 
$\frac { {\bf h}_p^T{\bf h}_q }{\Vert {\bf h}_p \Vert^2 + \Vert {\bf h}_q \Vert^2 }$ as
\begin{equation}
\label{l4ss1}
\frac{{\bf h}_p^T{\bf h}_q/(2N_t)}{(\Vert{\bf h}_p \Vert^2 + \Vert {\bf h}_q \Vert^2)/(2N_t)}.
\end{equation}
As $N_t \rightarrow \infty$, by strong law of large numbers, the 
denominator of (\ref{l4ss1}) converges to 1 almost surely. Also, the
numerator of (\ref{l4ss1}) can be written as
\begin{eqnarray}
\label{l4St2}
\frac {{\bf h}_p^T{\bf h}_q} {2N_t} & = & \frac {\sum_{k=1}^{2N_t} {h}_{p,k} {h}_{q,k}}{2N_t},
\end{eqnarray}
where $h_{p,k}$ and $h_{q,k}$ refer to the $k$th entry of the vectors 
${\bf h}_p$ and ${\bf h}_q$, respectively. Each ${h}_{p,k}{h}_{q,k}$ term 
in the summation in (\ref{l4St2}) has the same distribution and 
has mean 0. Therefore, by strong law of large numbers, we can see that 
$\frac{{\bf h}_p^T{\bf h}_q}{2N_t}$ converges to 0 almost surely.
This also implies that $\frac {{\bf h}_p^T{\bf h}_q} {2N_t}$ converges 
in distribution to the constant 0, and hence by Slutsky's theorem,
$\frac{{\bf h}_p^T{\bf h}_q}{\Vert{\bf h}_p\Vert^2 + \Vert{\bf h}_q\Vert^2}$ 
converges in distribution to 0. Since, if a sequence of r.v's converges in 
distribution to a constant then the sequence converges in probability to 
that constant, we conclude that indeed 
$\frac{{\bf h}_p^T{\bf h}_q}{\Vert{\bf h}_p\Vert^2 + \Vert{\bf h}_q\Vert^2}$ 
converges in probability to 0. This proves the the base case.

{\em Induction Hypothesis:} Let $z_{{\bf u}_n,{\bf d}} \inprob 0$
as $N_t\rightarrow \infty$, $\forall \, n=2,3,\cdots,m$.

\vspace{-2mm}
{\em Induction Step:}
Proof for $n=m+1$: We have
\begin{eqnarray}
\label{l4St3}
z_{{\bf u}_{(m+1)},{\bf d}} &=& \frac {\sum_{k=1}^{m+1} \sum_{j=k+1}^{m+1} {\bf h}_{i_j}^T{\bf h}_{i_k} d_{i_j}d_{i_k}}{\sum_{j=1}^{m+1} \Vert {\bf h}_{i_j}\Vert^2} \\ 
& = & \frac{ \sum_{k=1}^{m} \sum_{j=k+1}^{m} {\bf h}_{i_j}^T{\bf h}_{i_k} d_{i_j}d_{i_k} + \sum_{k = 1}^{m} {\bf h}_{i_{(m+1)}}^T{\bf h}_{i_k}d_{i_{(m+1)}} d_{i_k}} {\Vert {\bf h}_{i_{(m+1)}} \Vert^2 + \sum_{j=1}^{m} 
\Vert {\bf h}_{i_j}\Vert^2 } \\ \label{new1}
& = & \frac {\frac {
\sum_{k=1}^{m} \sum_{j=k+1}^{m} {\bf h}_{i_j}^T{\bf h}_{i_k} d_{i_j}d_{i_k} 
} {\sum_{j=1}^{m} \Vert {\bf h}_{i_j}\Vert^2 } + \frac {\sum_{k = 1}^{m} {\bf h}_{i_{(m+1)}}^T{\bf h}_{i_k}d_{i_{(m+1)}} d_{i_k}} {\sum_{j=1}^{m} \Vert {\bf h}_{i_j}\Vert^2 } } { 1 + \frac {\Vert {\bf h}_{i_{(m+1)}} \Vert^2 } { \sum_{j=1}^{m} \Vert {\bf h}_{i_j}\Vert^2 } }.
\end{eqnarray}
Using Slutsky's theorem and the strong law of large numbers, it can be 
shown that the denominator in (\ref{new1}) converges to $(1+\frac{1}{m})$ 
in probability. Also, from the induction hypothesis, the term 
$\frac{\sum_{k=1}^{m} \sum_{j=k+1}^{m} {\bf h}_{i_j}^T{\bf h}_{i_k} d_{i_j}d_{i_k}}{\sum_{j=1}^{m} \Vert {\bf h}_{i_j}\Vert^2}$
in the numerator of (\ref{new1}) converges in probability to 0.
Therefore, the numerator in (\ref{new1}) converges to the same 
distribution that the term 
$\frac{\sum_{k=1}^{m}{\bf h}_{i_{(m+1)}}^T{\bf h}_{i_k}d_{i_{(m+1)}}d_{i_k}}{\sum_{j=1}^{m} \Vert{\bf h}_{i_j}\Vert^2}$ 
converges to. Also, the term 
$\frac{\sum_{k=1}^{m}{\bf h}_{i_{(m+1)}}^T{\bf h}_{i_k}d_{i_{(m+1)}}d_{i_k}}{\sum_{j=1}^{m} \Vert{\bf h}_{i_j}\Vert^2}$ 
is the same as 
$\frac{(\sum_{k=1}^{m}{\bf h}_{i_{(m+1)}}^T{\bf h}_{i_k}d_{i_{(m+1)}}d_{i_k})/(mN_t)}{(\sum_{j=1}^{m} \Vert{\bf h}_{i_j}\Vert^2)/(mN_t)}$.
Further, from the strong law of large numbers, the term
$(\sum_{j=1}^{m} \Vert{\bf h}_{i_j}\Vert^2)/(mN_t)$ converges almost surely 
to 1. Therefore, from Slutsky's theorem, we know that 
$\frac{(\sum_{k=1}^{m}{\bf h}_{i_{(m+1)}}^T{\bf h}_{i_k}d_{i_{(m+1)}}d_{i_k})/(mN_t)}{(\sum_{j=1}^{m} \Vert {\bf h}_{i_j}\Vert^2)/(mN_t)}$ 
converges in distribution to the distribution to which the term
$(\sum_{k=1}^{m}{\bf h}_{i_{(m+1)}}^T{\bf h}_{i_k}d_{i_{(m+1)}}d_{i_k})/(mN_t)$
converges.

For a given vector ${\bf d}$, ${\bf h}_{i_k}d_{i_k}$ is a random vector 
whose distribution is the same as that of ${\bf h}_{i_k}$. Therefore, 
applying Lemma \ref{lem4}, we see that the term 
$(\sum_{k=1}^{m}{\bf h}_{i_{(m+1)}}^T{\bf h}_{i_k}d_{i_{(m+1)}}d_{i_k})/(mN_t)$
converges almost surely to 0. Hence, the numerator in (\ref{new1}) converges 
in probability to the constant 0 . Therefore, 
$z_{{\bf u}_{(m+1)},{\bf d}} \inprob 0$
as $N_t\rightarrow \infty$. This proves the induction step and completes the 
proof of Lemma \ref{lem5}.
$\square$

\vspace{-4mm}
\section*{Appendix B: Proof of Theorem \ref{thm1}}
\vspace{-5mm}
We present the proof of Theorem \ref{thm1} in this appendix.
We shall prove through induction that if
${\bf n} \in {\cal R}_{{\bf d}^{1}}$, then
${\bf n} \in {\cal R}_{{\bf d}^{m}}$ in probability,
$\forall m=2,\cdots,2N_t$, as $N_t \rightarrow \infty$.
{\em Base Case ($m=2$):}
Let ${\bf n} \in {\cal R}_{{\bf d}^{1}}$. Therefore, from the
definition of ${\cal R}_{{\bf d}^{m}}$, ${\bf n}$ satisfies
(\ref{l1p5}). We show that
${\bf n} \in {\cal R}_{{\bf d}^{2}}$ in probability as
$N_t \rightarrow \infty$. For ${\bf n}$ to belong to
${\cal R}_{{\bf d}^{2}}$, in addition to satisfying (\ref{l1p5}),
${\bf n}$ must also satisfy the following equation
$\forall \, p,q = 1,\cdots 2N_t, p \neq q$:
\begin{eqnarray}
\vspace{-2mm}
\label{L2}
\big({\bf n} + {\bf H}({\bf x} - {\bf d}) + {\bf h}_{p}{d}_{p} + {\bf h}_{q}{d}_{q} \big)^T \big({\bf h}_{p}{d}_{p} + {\bf h}_{q}{d}_{q} \big) & \geq & 0,
\end{eqnarray}
which can be rewritten as
\begin{eqnarray}
\label{L21}
\vspace{-2mm}
\big({\bf n} + {\bf H}({\bf x} - {\bf d})\big)^T{\bf h}_{p}{d}_{p} + \big( {\bf n} + {\bf H}({\bf x} - {\bf d})\big)^T{\bf h}_{q}{d}_{q}  & 
\geq &  -\Vert {\bf h}_{p} \Vert^2 - \Vert {\bf h}_{q} \Vert^2 -2{d}_{p}{d}_{q}{\bf h}_{p}^T{\bf h}_{q}.
\end{eqnarray}
Since ${\bf n}$ satisfies (\ref{l1p5}), it satisfies the following two
equations:
\vspace{-2mm}
\begin{eqnarray}
\label{L22}
\nonumber
\big({\bf n} + {\bf H}({\bf x} - {\bf d})\big)^T {\bf h}_{p}{d}_{p} & \geq & -\Vert {\bf h}_{p} \Vert^2, \\
\big({\bf n} + {\bf H}({\bf x} - {\bf d})\big)^T {\bf h}_{q}{d}_{q} & \geq & -\Vert {\bf h}_{q} \Vert^2.
\end{eqnarray}

\vspace{-3mm}
Comparing (\ref{L22}) and (\ref{L21}), we notice that if ${\bf h}_{p}$
and ${\bf h}_{q}$ are orthogonal, then ${\bf n}$ trivially satisfies
(\ref{L21}) for all $N_t$. Therefore, when ${\bf h}_{p}$ and ${\bf h}_{q}$
are non-orthogonal, the only extra term in the RHS of (\ref{L21}) is
$2{d}_{p}{d}_{q}{\bf h}_{p}^T{\bf h}_{q}$. Applying Lemma \ref{lem5}, 
with $n=2$, we see that as $N_t \rightarrow \infty$, the r.v.
$\frac{{\bf h}_{p}^T{\bf h}_{q}}{\Vert {\bf h}_{p} \Vert^2 + \Vert {\bf h}_{q} \Vert^2}$
converges to zero in probability. 
Then, we can write, for any $\epsilon$, $0 \leq \epsilon \leq 1$
\begin{eqnarray}
\label{one}
p\left(\frac{|{\bf h}_{p}^T{\bf h}_{q}| }   
{\Vert{\bf h}_{p} \Vert^2 + \Vert {\bf h}_{q} \Vert^2} > \epsilon
\right) & < & \epsilon, \,\,\, \forall N_t>f(\epsilon).
\end{eqnarray}
Now, let us analyze $p({\bf n} \in {\cal R}_{{\bf d}^{2}})$
for the case of $d_pd_q=+1$ (a similar analysis holds for $d_pd_q=-1$).
Consider two disjoint events 
$E_1 = \left\{\frac{|{\bf h}_{p}^T{\bf h}_{q}|}{\Vert{\bf h}_{p} \Vert^2 + \Vert{\bf h}_{q} \Vert^2} < \epsilon\right\}$ and
$E_2 = \left\{\frac{|{\bf h}_{p}^T{\bf h}_{q}|}{\Vert{\bf h}_{p} \Vert^2 + \Vert{\bf h}_{q} \Vert^2} > \epsilon\right\}$. Then, we can write
\begin{eqnarray}
\label{two}
p({\bf n} \notin {\cal R}_{{\bf d}^{2}}) & = &
p({\bf n} \notin {\cal R}_{{\bf d}^{2}}|E_1) \, p(E_1) \, + \, 
p({\bf n} \notin {\cal R}_{{\bf d}^{2}}|E_2) \, p(E_2). 
\end{eqnarray}
The event $E_1$ can be further split into two disjoint events 
$E_{11}$ and $E_{12}$, given by
$E_{11} = \left\{ 0 < {\bf h}_p^T {\bf h}_q < \epsilon \left(\Vert{\bf h}_{p} \Vert^2 + \Vert{\bf h}_{q}\Vert^2\right) \right\}$
and
$E_{12} = \left\{ 0 > {\bf h}_p^T {\bf h}_q > -\epsilon \left(\Vert{\bf h}_{p} \Vert^2 + \Vert{\bf h}_{q}\Vert^2\right) \right\}$.
Also, from (\ref{one}), $p(E_1)>1-\epsilon$ and $p(E_2) < \epsilon$.
Therefore, using (\ref{two}), we can write 
\begin{eqnarray}
\nonumber
\label{five}
p({\bf n} \notin {\cal R}_{{\bf d}^{2}}) & < &
p({\bf n} \notin {\cal R}_{{\bf d}^{2}}|E_1) \, p(E_1) \, + \, \epsilon \\ \nonumber
& < & p({\bf n} \notin {\cal R}_{{\bf d}^{2}}|E_{11}) \, p(E_{11}) \, + \, p({\bf n} \notin {\cal R}_{{\bf d}^{2}}|E_{12}) \, p(E_{12}) \, + \, \epsilon \\ 
& < & p({\bf n} \notin {\cal R}_{{\bf d}^{2}}|E_{11}) \, + \, p({\bf n} \notin {\cal R}_{{\bf d}^{2}}|E_{12}) \, + \, \epsilon. 
\end{eqnarray}
If event $E_{11}$ is true, then
\begin{eqnarray}
\label{three}
\hspace{-8mm}
-\left(\Vert{\bf h}_{p} \Vert^2 + \Vert{\bf h}_{q}\Vert^2\right) & > &
-\left(\Vert{\bf h}_{p} \Vert^2 + \Vert{\bf h}_{q}\Vert^2 + 2{\bf h}_p^T {\bf h}_q \right) \,\,\, > \,\,\, 
-\left(\Vert{\bf h}_{p} \Vert^2 + \Vert{\bf h}_{q}\Vert^2\right) \left(1+2\epsilon\right).
\end{eqnarray}
Since ${\bf n} \in {\cal R}_{{\bf d}^{1}}$, ${\bf n}$ satisfies 
(\ref{L22}), and hence satisfies the following equation:
\begin{eqnarray}
\label{four}
\left({\bf n} + {\bf H}({\bf x}-{\bf d})\right)^T \left({\bf h}_pd_p + {\bf h}_qd_q \right) & \geq & -\left(\Vert{\bf h}_{p} \Vert^2 + \Vert{\bf h}_{q}\Vert^2\right).  
\end{eqnarray}
Using (\ref{three}) and (\ref{four}), we see that ${\bf n}$ satisfies
(\ref{L21}), and therefore ${\bf n} \in {\cal R}_{{\bf d}^{2}}$.
Hence, we can conclude that 
$p\left({\bf n} \notin {\cal R}_{{\bf d}^{2}}|E_{11}\right) = 0$.
Now, we can rewrite (\ref{five}) as
\begin{eqnarray}
\label{seven}
p({\bf n} \notin {\cal R}_{{\bf d}^{2}}) & < &  p({\bf n} \notin {\cal R}_{{\bf d}^{2}}|E_{12}) \, + \, \epsilon.
\end{eqnarray}
If event $E_{12}$ is true, then
\begin{eqnarray}
\label{six}
\hspace{-8mm}
-\left(\Vert{\bf h}_{p} \Vert^2 + \Vert{\bf h}_{q}\Vert^2\right) (1-2\epsilon) & > & -\left(\Vert{\bf h}_{p} \Vert^2 + \Vert{\bf h}_{q}\Vert^2 + 2{\bf h}_p^T {\bf h}_q \right) \,\,\, > \,\,\,
-\left(\Vert{\bf h}_{p} \Vert^2 + \Vert{\bf h}_{q}\Vert^2\right).
\end{eqnarray}
Using (\ref{L21}) and (\ref{six}), we can write that 

\vspace{-4mm}
{\footnotesize 
\begin{eqnarray}
\nonumber
\hspace{-0mm}
p({\bf n} \notin {\cal R}_{{\bf d}^{2}} \, | \, E_{12}) & = & p\left( \big[   
\big({\bf n} + {\bf H}({\bf x} - {\bf d})\big)^T{\bf h}_{p}{d}_{p} + \big( {\bf n} + {\bf H}({\bf x} - {\bf d})\big)^T{\bf h}_{q}{d}_{q} \,\, 
\leq \,\,  -\Vert {\bf h}_{p} \Vert^2 - \Vert {\bf h}_{q} \Vert^2 -2{\bf h}_{p}^T{\bf h}_{q}\big] \, \big| \, E_{12}  \right) \nonumber \\
\hspace{-0cm} & \hspace{-3.7cm} < & \hspace{-2.0cm}
p\left(\big[-\left(\Vert{\bf h}_{p} \Vert^2 + \Vert{\bf h}_{q}\Vert^2\right) \, \leq \,
\left({\bf n} + {\bf H}({\bf x}-{\bf d})\right)^T \left({\bf h}_pd_p + {\bf h}_qd_q \right) 
\, \leq \, -\left(\Vert{\bf h}_{p} \Vert^2 + \Vert{\bf h}_{q}\Vert^2\right)(1-2\epsilon) \big] \, \big| \, E_{12} \right)\hspace{-1mm}. 
\label{eqxx}
\end{eqnarray}
}

\vspace{-8mm}
Define ${\cal R}_{\epsilon}$ to be a set of vectors in ${\mathbb R}^{2N_t}$, 
as 
{\footnotesize
\begin{eqnarray}
\hspace{-6mm}
{\cal R}_{\epsilon} & \Define & 
\left\{{\bf v} \,\, \big| \,\, -\left(\Vert{\bf h}_{p} \Vert^2 + \Vert{\bf h}_{q}\Vert^2\right) \, \leq
\left({\bf v} + {\bf H}({\bf x}-{\bf d})\right)^T \left({\bf h}_pd_p + {\bf h}_qd_q \right) 
\, \leq \, -\left(\Vert{\bf h}_{p} \Vert^2 + \Vert{\bf h}_{q}\Vert^2\right)(1-2\epsilon)\right\}.
\label{eqxx2}
\end{eqnarray}
}
Also, define a function $f_2$ as
\begin{eqnarray}
f_2(\epsilon)& \Define & p({\bf n} \in {\cal R}_{\epsilon} \, | \, E_{12}).
\end{eqnarray}
Using the above definitions, (\ref{eqxx}) can rewritten as
\begin{eqnarray}
p({\bf n} \notin {\cal R}_{{\bf d}^{2}} \, | \, E_{12}) & < & f_2(\epsilon). 
\label{eqxx3}
\end{eqnarray}
Let $\epsilon_1, \epsilon_2 \in {\mathbb R}$, 
$\epsilon_1, \epsilon_2 > 0$, and $\epsilon_1>\epsilon_2$.
From the definition of ${\cal R}_{\epsilon}$ in (\ref{eqxx2}), 
it can be seen that ${\cal R}_{\epsilon_2} \subset {\cal R}_{\epsilon_1}$.
This implies that $f_2(\epsilon_1) > f_2(\epsilon_2)$. Hence $f_2$ is a
monotonically increasing function. Using (\ref{eqxx3}), 
we can rewrite (\ref{seven}) as
written as
\begin{eqnarray}
p({\bf n} \notin {\cal R}_{{\bf d}^{2}}) & < & f_2(\epsilon) \,+\, \epsilon.
\end{eqnarray}
Therefore,
\begin{eqnarray}
\label{eight}
p({\bf n} \in {\cal R}_{{\bf d}^{2}}) & > & 1-\left(f_2(\epsilon) \,+\, \epsilon\right).
\end{eqnarray}
Now define $g_2(\epsilon) \Define f_2(\epsilon) + \epsilon$. So 
$g_2$ is a monotonic function and is therefore invertible.
Let $\delta=g_2(\epsilon)$. Using (\ref{one}) and the above definitions, we
can write that 
\begin{eqnarray}
N_t & > & f(\epsilon) \nonumber \\
& > & f\left(g_2^{-1}(\delta)\right) \nonumber \\
& > & N_2(\delta),
\end{eqnarray}
where $N_2 \Define f \circ g_2^{-1}$. We can then write (\ref{eight}) as
\begin{eqnarray}
p({\bf n} \in {\cal R}_{{\bf d}^{2}}) & > & 1 - \delta.
\end{eqnarray}
Since $g_2$ is a continuous monotonic function,
for any
$\delta$, $0 \leq \delta \leq 1$, there exists an integer $N_2(\delta)$
such that for $N_t > N_2(\delta)$, $p({\bf n} \in {\cal R}_{{\bf d}^{2}})
> 1-\delta$.
Therefore,
${\bf n} \in {\cal R}_{{\bf d}^{2}}$ in probability as
$N_t \rightarrow \infty$, thus proving the base case.

\vspace{-2mm}
{\em Induction Hypothesis:}
Let ${\bf n} \in {\cal R}_{{\bf d}^{m-1}}$ in probability as
$N_t \hspace{-1mm} \rightarrow \hspace{-1mm} \infty$. 

\vspace{-2mm}
{\em Induction Step:}
We need to prove that
${\bf n} \in {\cal R}_{{\bf d}^{m}}$ in probability as
{\small $N_t \rightarrow \infty$}. For ${\bf n}$ to belong
to  ${\cal R}_{{\bf d}^{m}}$, ${\bf n}$ must satisfy the following
equation for all possible $m$-tuples $(i_1,i_2,\cdots,i_m)$:

\vspace{-12mm}
\begin{eqnarray}
\label{Lt2}
\Big({\bf n} + {\bf H}({\bf x} - {\bf d}) + \Big(\sum_{j=1}^{m} {\bf h}_{i_{j}}{d}_{i_{j}}\Big)\Big)^T \bigg(\sum_{j=1}^{m} {\bf h}_{i_{j}}{d}_{i_{j}}\bigg) & \geq & 0,
\end{eqnarray}

\vspace{-4mm}
which can be written as

\vspace{-12mm}
\begin{eqnarray}
\label{Lt2x}
\nonumber
\big({\bf n} + {\bf H}({\bf x} - {\bf d})\big)^T \Big(\sum_{j=1}^{m-1} {\bf h}_{i_{j}}{d}_{i_{j}}\Big) +
\big({\bf n} + {\bf H}({\bf x} - {\bf d})\big)^T {\bf h}_{i_{m}}{d}_{i_{m}} &  \\
& 
\hspace{-70mm}
\geq \,\,  -\Vert  \sum_{j = 1}^{m-1} {\bf h}_{i_{j}}{d}_{i_{j}} \Vert^2  - \Vert {\bf h}_{i_{m}} \Vert^2 - 2\Big(\sum_{j = 1}^{m-1} {\bf h}_{i_{j}}{d}_{i_{j}}\Big)^T{\bf h}_{i_{m}}{d}_{i_{m}}.
\end{eqnarray}

\vspace{-4mm}
However, we know from the induction hypothesis that
$({\bf n} + {\bf H}({\bf x} - {\bf d}))^T (\sum_{j = 1}^{m-1} {\bf h}_{i_{j}}{d}_{i_{j}}) \geq -\Vert  \sum_{j = 1}^{m-1} {\bf h}_{i_{j}}{d}_{i_{j}} \Vert^2$.
Also, since
${\bf n} \in {\cal R}_{{\bf d}^1}$, we know that $( {\bf n} + {\bf H}({\bf x} - {\bf d}))^T {\bf h}_{i_{m}}{d}_{i_{m}} \geq -\Vert {\bf h}_{i_{m}} \Vert^2 $.
Therefore, if the term
$2(\sum_{j=1}^{m-1} {\bf h}_{i_{j}}{d}_{i_{j}})^T{\bf h}_{i_{m}}{d}_{i_{m}}$
in the RHS of (\ref{Lt2x}) were 0, then (\ref{Lt2}) would have been
trivially satisfied.
We now show that the contribution of the term
$2(\sum_{j=1}^{m-1} {\bf h}_{i_{j}}{d}_{i_{j}})^T{\bf h}_{i_{m}}{d}_{i_{m}}$
when compared to the other two terms in the RHS (\ref{Lt2x}) converges to 0
as $N_t \rightarrow \infty$.

Define a r.v.
$v_m \, \Define \, \frac {2(\sum_{j=1}^{m-1} {\bf h}_{i_{j}}{d}_{i_{j}})^T{\bf h}_{i_{m}}{d}_{i_{m}}}{\Vert \sum_{j=1}^{m-1} {\bf h}_{i_{j}}{d}_{i_{j}} \Vert^2 + \Vert {\bf h}_{i_{m}} \Vert^2 }$.
Our objective is to show that as $N_t \rightarrow \infty$,
$v_m \rightarrow 0$ in probability.
This is equivalent to proving that
$w_m \Define v_m + 1 = \frac{\Vert \sum_{j = 1}^{m}{\bf h}_{i_{j}}{d}_{i_{j}} \Vert^2}{\Vert {\bf h}_{i_{m}} \Vert^2 + \Vert \sum_{j = 1}^{m-1} {\bf h}_{i_{j}}{d}_{i_{j}} \Vert^2 }$
converges to one in probability as $N_t \rightarrow \infty$.
We can write $w_m$ as
\begin{eqnarray}
\label{Lt3}
w_m & = & \frac{\frac{\Vert \sum_{j = 1}^{m} {\bf h}_{i_{j}}{d}_{i_{j}} \Vert^2}{\sum_{j = 1}^{m} \Vert {\bf h}_{i_{j}}\Vert^2}}{\frac{\Vert {\bf h}_{i_{m}} \Vert^2 } {  \sum_{j=1}^{m} \Vert {\bf h}_{i_{j}}\Vert^2} + \frac {\Vert (\sum_{j = 1}^{m-1} {\bf h}_{i_{j}}{d}_{i_{j}} ) \Vert^2}{\sum_{j = 1}^{m} \Vert {\bf h}_{i_{j}}\Vert^2} }.
\end{eqnarray}
From Lemma \ref{lem5}, we know that for any integer $m$, $1\leq m \leq 2N_t$,
it is true that
$\frac{\sum_{k=1}^{m} \sum_{j=k+1}^{m} {\bf h}_{i_j}^T{\bf h}_{i_k} d_{i_j}d_{i_k}} { \sum_{j=1}^{m} \Vert {\bf h}_{i_j}\Vert^2} $
converges to 0 in probability as $N_t \rightarrow \infty$.
By Slutsky's theorem, this is equivalent to

\vspace{-8mm}
\begin{eqnarray}
\label{Lt5}
\frac{2\sum_{k=1}^{m} \sum_{j=k+1}^{m} {\bf h}_{i_j}^T{\bf h}_{i_k} d_{i_j}d_{i_k} } { \sum_{j=1}^{m} \Vert {\bf h}_{i_j} \Vert^2 }  + 1 & = &
\frac{ \Vert \sum_{j = 1}^{j = m} {\bf h}_{i_{j}}{d}_{i_{j}} \Vert^2 } { \sum_{j = 1}^{m} \Vert {\bf h}_{i_{j}}\Vert^2 } \,\, \inprob \,\, 1
\end{eqnarray}

\vspace{-2mm}
as $N_t \rightarrow \infty$. We shall use this result to prove the
convergence of $w_m$ in (\ref{Lt3}). Using (\ref{Lt5}), it can be seen
that the numerator of $w_m$ in (\ref{Lt3}) converges to 1 as $N_t\rightarrow
\infty$, i.e.,
\begin{eqnarray}
\label{Lt9}
\frac{\Vert \sum_{j = 1}^{m} {\bf h}_{i_{j}}{d}_{i_{j}} \Vert^2} {\sum_{j = 1}^{m} \Vert {\bf h}_{i_{j}}\Vert^2} & \inprob & 1, \,\,\,\ \mbox{as} \,\, N_t \rightarrow \infty.
\end{eqnarray}
In the denominator of (\ref{Lt3}), it can be shown that the term
\begin{eqnarray}
\label{Lt10}
\frac{\Vert{\bf h}_{i_{m}} \Vert^2}{\sum_{j=1}^{m} \Vert{\bf h}_{i_{j}}\Vert^2} & \inprob & \frac{1}{m}, \,\,\,\ \mbox{as} \,\, N_t \rightarrow \infty.
\end{eqnarray}
The 2nd term in the denominator of (\ref{Lt3}) can be rewritten as
\begin{eqnarray}
\label{Lt7}
\frac{\Vert(\sum_{j = 1}^{m-1} {\bf h}_{i_{j}}{d}_{i_{j}}) \Vert^2}{\sum_{j=1}^{m} \Vert {\bf h}_{i_{j}}\Vert^2} &
= & \frac { \frac {\Vert  (\sum_{j = 1}^{m-1} {\bf h}_{i_{j}}{d}_{i_{j}}) \Vert^2   } { \sum_{j = 1}^{m-1} \Vert {\bf h}_{i_{j}}\Vert^2 }}{\frac {\Vert {\bf h}_{i_{m}}\Vert^2  } { \sum_{j = 1}^{m-1} \Vert {\bf h}_{i_{j}}\Vert^2  }   +  1  }.
\end{eqnarray}
Similar to the derivation of (\ref{Lt5}), 
we can claim that the numerator in (\ref{Lt7})
converges to one in probability. From Slutsky's theorem, it can be shown that
$\frac{\Vert{\bf h}_{i_{m}}\Vert^2}{\sum_{j=1}^{m-1} \Vert{\bf h}_{i_{j}}\Vert^2}$
converges to $\frac{1} {m-1}$ in probability.
Using this and Slutsky's theorem, it can be shown that (\ref{Lt7})
converges to $\frac { m-1 } { m}$ in probability.
Using this result along with (\ref{Lt9}),(\ref{Lt10}) and Slutsky's
theorem in (\ref{Lt3}), it can be shown that $w_m$ converges to one
in probability as $N_t \rightarrow \infty$.
This, therefore, implies that $v_m$ converges to zero in probability.
As proved in the base case,
it can be shown that for any $\delta$, $0 \leq \delta \leq 1$, there
exists an integer $N_m(\delta)$ such that for $N_t > N_m(\delta)$,
$p({\bf n} \in {\cal R}_{{\bf d}^{m}}) > 1-\delta$. This proves the
induction step and completes the proof of Theorem \ref{thm1}.
$\square$

\vspace{-6.0mm}
{\footnotesize 
\bibliographystyle{IEEE}

}

\newpage

\begin{figure}
\begin{center}
\epsfxsize=10.5cm
\epsfbox{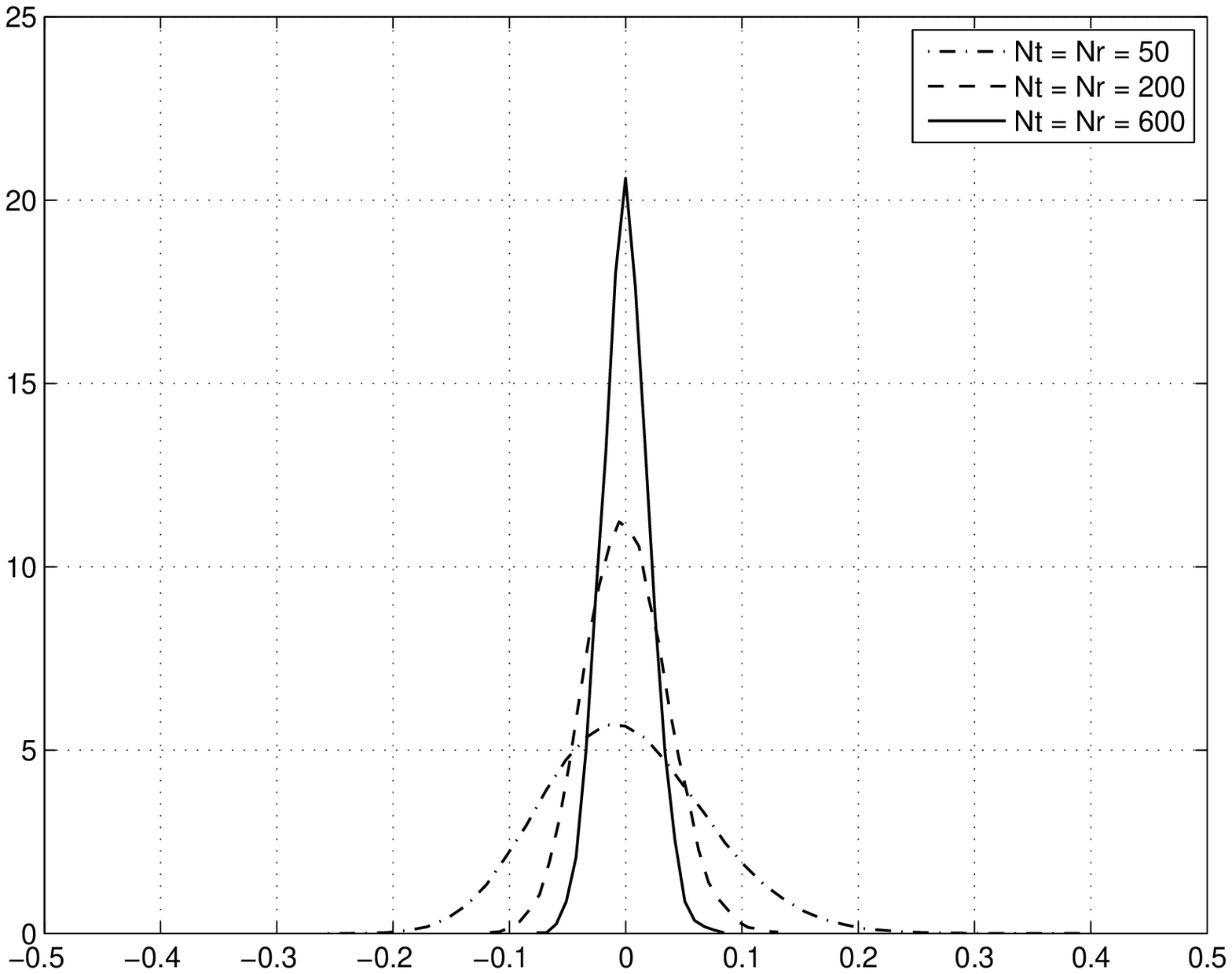}
\caption{Simulated pdf of $z_{{\bf u}_n,{\bf d}}$ for 
$n=2N_t$ for increasing $N_t=N_r$. 4-QAM. The pdf tends towards Dirac
delta function at zero.} 
\label{fig0}
\end{center}
\end{figure}

\begin{figure}
\begin{center}
\epsfxsize=10.5cm
\epsfbox{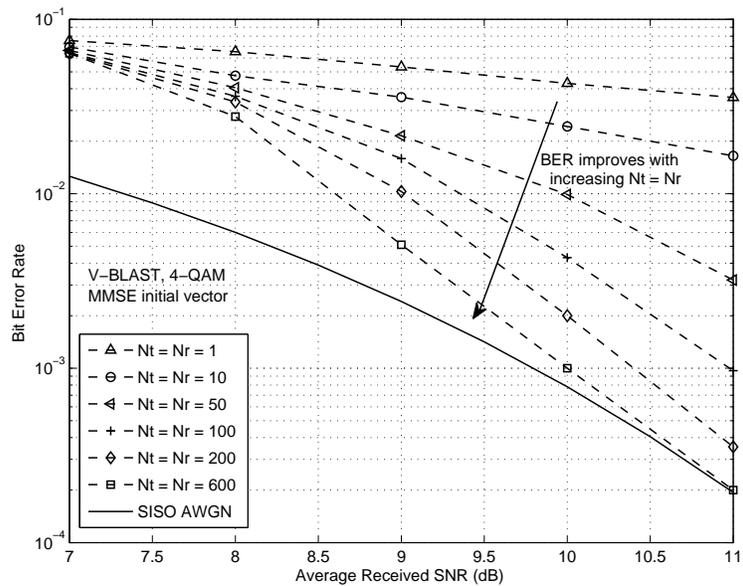}
\caption{Simulated BER performance of the LAS detector for V-BLAST as a 
function of average received SNR for increasing values of $N_t=N_r$. 
MMSE initial vector, {\bf 4-QAM}. LAS detector achieves near SISO AWGN 
performance at high SNRs for large $N_t=N_r$.}
\label{fig1}
\end{center}
\end{figure}

\begin{figure}
\begin{center}
\epsfxsize=10.5cm
\epsfbox{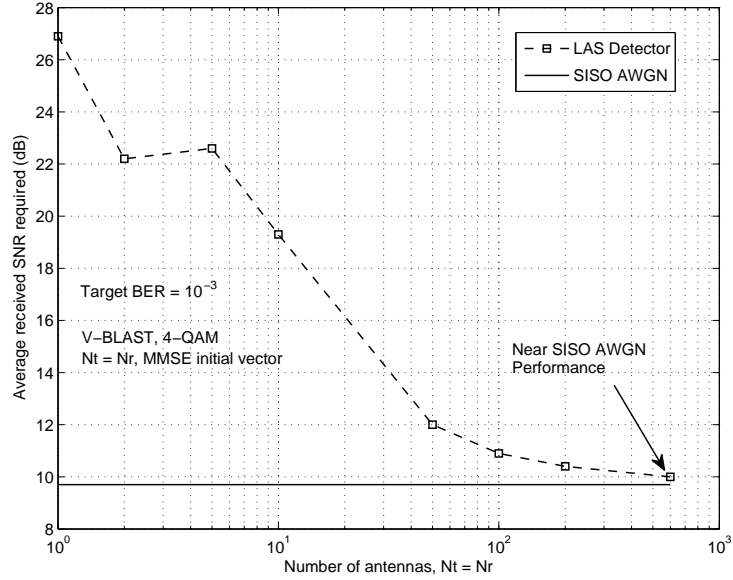}
\caption{Average received SNR required to achieve a target BER of 
$10^{-3}$ in V-BLAST for increasing values of $N_t=N_r$ for {\bf 4-QAM}. 
LAS detector with MMSE initial vector. LAS detector achieves near SISO 
AWGN performance for large $N_t=N_r$.}
\label{fig2}
\end{center}
\end{figure}

\begin{figure}
\begin{center}
\epsfxsize=10.5cm
\epsfbox{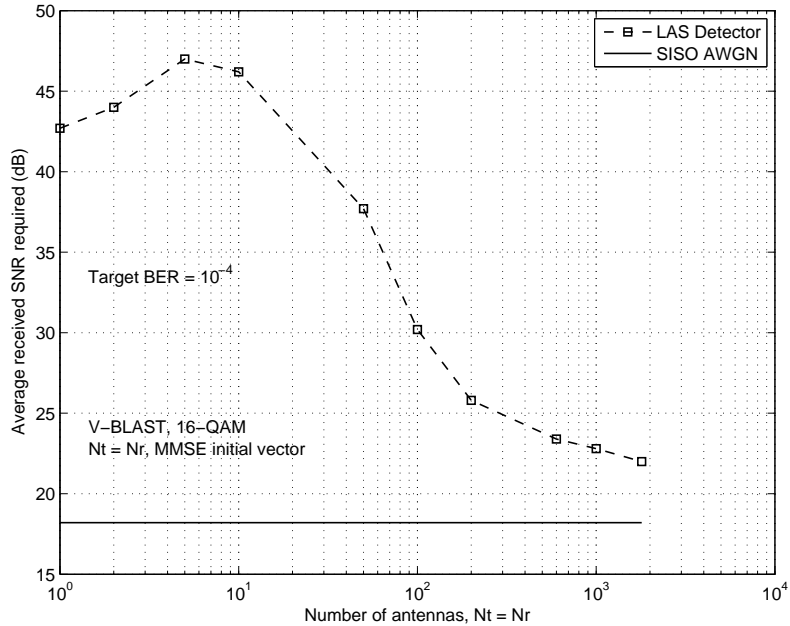}
\caption{Average received SNR required to achieve a target BER
of $10^{-4}$ in V-BLAST for increasing values of $N_t=N_r$ for {\bf 16-QAM}. 
LAS detector with MMSE initial vector. LAS detector performance approaches
SISO AWGN performance for large $N_t=N_r$.}
\label{fig3}
\end{center}
\end{figure}

\end{document}